%% file: main.tex
\def\jcap{JCAP}
\def\mnras{MNRAS}             
\def\aap{A\&A}             
\def\apjs{ApJS}             
\def\apjl{ApJL}             

\def\physrep{ßPhys.~Rep.}   

\documentclass[%
reprint,
superscriptaddress,
nofootinbib,
nolongbibliography,
 amsmath,amssymb,
 aps,
prd,
floatfix,
]{revtex4-2}

\usepackage{graphicx}
\usepackage{dcolumn}
\usepackage{bm}
\usepackage{xcolor}

\definecolor{maroon}{RGB}{192,0,0}
\usepackage[colorlinks=true, citecolor=maroon,
linkcolor=maroon,
urlcolor=blue]{hyperref}
\usepackage{times}
\usepackage{tikz}

\usepackage{amsmath}	
\usepackage{amssymb}    
\usepackage{amssymb}	
\usepackage{hyperref}
\usepackage{contour}
\usepackage{ulem}
\usepackage{float}
\usepackage{pgfplots}
\pgfplotsset{compat=1.18}
\usepackage{enumitem}

\newcommand{\rockstar}{\textsc{Rockstar}}

\newcommand{\gadget}{\textsc{GADGET-2}}

\newcommand{\hmsun}{h^{-1}\ {\rm M_{\odot}}}
\newcommand{\hkpc}{h^{-1}\ {\rm kpc}}
\newcommand{\hMpc}{h^{-1}\ {\rm Mpc}}
\newcommand{\hMpcinv}{h\ {\rm Mpc}^{-1}}
\newcommand{\hGpc}{h^{-1}\ {\rm Gpc}}

\newcommand{\ximm}{\xi_{\rm mm}}
\newcommand{\xihm}{\xi_{\rm hm}}
\newcommand{\xiinf}{\xi_{\rm inf}}
\newcommand{\xilin}{\xi_{\rm lin}}
\newcommand{\xizel}{\xi_{\rm ZA}}

\newcommand{\rh}{r_{\rm h}}
\newcommand{\rinf}{r_{\rm inf}}

\newcommand{\Morb}{M_{\rm orb}}

\newcommand{\avg}[1]{\langle #1 \rangle}

\newcommand{\rhoorb}{\rho_{\rm orb}}
\newcommand{\rhoinf}{\rho_{\rm inf}}
\newcommand{\barrhom}{{\bar \rho}_{\rm m}}

\newcommand{\aacc}{a_{\rm acc}}

\newcommand{\hMsun}{h^{-1}\ M_\odot}
\newcommand{\kNL}{k_{\rm NL}}
\newcommand{\xilarge}{\xi_{\rm large}}

\begin{document}

\title[Dynamics-Based Halo Model]{Dynamics-based halo model for large scale structure}


\author{Edgar M. Salazar}
    \email{edgarmsc@arizona.edu}
    \affiliation{Department of Physics, University of Arizona, Tucson, AZ 85721, USA}
\author{Eduardo Rozo}
    \affiliation{Department of Physics, University of Arizona, Tucson, AZ 85721, USA}
\author{Rafael Garc\'ia}
    \affiliation{Department of Physics, University of Arizona, Tucson, AZ 85721, USA}
\author{Nickolas Kokron}
    \affiliation{Department of Astrophysical Sciences, Princeton University, Princeton, NJ, 08540, USA}
    \affiliation{School of Natural Sciences, Institute for Advanced Study, Princeton, NJ, 08540, USA}
\author{Susmita Adhikari}
    \affiliation{Department of Physics, Indian Institute of Science Education and Research, Homi Bhaba Road, Pashan, Pune 411008, India}
\author{Benedikt Diemer}
    \affiliation{Department of Astronomy, University of Maryland, College Park, MD 20742, USA}
\author{Calvin Osinga}
    \affiliation{Department of Astronomy, University of Maryland, College Park, MD 20742, USA}

\date{\today}

\begin{abstract}
Accurate modelling of the one-to-two halo transition has long been difficult to achieve.  We demonstrate that physically motivated halo definitions that respect the bimodal phase-space distribution of dark matter particles near halos resolves this difficulty.  Specifically, the two phase-space components are overlapping and correspond to: 1) particles \it orbiting \rm the halo; and 2) particles \it infalling \rm into the halo for the first time.  Motivated by this decomposition, Garc\'ia [R. Garc\'ia et. al., MNRAS 521, 2464 (2023)] 
advocated for defining haloes as the collection of particles orbiting their self-generated potential. This definition identifies the traditional one-halo term of the halo--mass correlation function with the distribution of orbiting particles around a halo, while the two-halo term governs the distribution of infalling particles.  We use dark matter simulations to demonstrate that the distribution of orbiting particles is finite and can be characterised by a single physical scale $\rh$, which we refer to as the \it halo radius. \rm   The two-halo term is described using a simple yet accurate empirical model based on the Zel'dovich correlation function. We further demonstrate that the halo radius imprints itself on the distribution of infalling particles at small scales.  Our final model for the halo--mass correlation function is accurate at the $\approx 2\%$ level for $r \in [0.1,50]\ \hMpc$.  The Fourier transform of our best fit model describes the halo--mass power spectrum with comparable accuracy for $k\in [0.06, 6.0]\ \hMpcinv$.
\end{abstract}

\maketitle


\input{sections/intro}
\input{sections/data}
\input{sections/orbiting}
\input{sections/infall}
\input{sections/halo_matter}
\input{sections/summary}

\section*{Data Availability}
The data underlying this article will be shared on reasonable request to the corresponding author.

\begin{acknowledgments}
E. Salazar is supported by NSF grant 3047980.  In addition to NSF 3047980, E. Rozo is also funded by DOE grant DE-SC0009913. NK acknowledges support from NSF award AST-2108126.
\end{acknowledgments}

\appendix
\input{sections/app_mass}

\bibliography{database}

\end{document}

%% file: sections/intro.tex
\section{Introduction}
\label{sec:intro}

The halo model is a well-established tool for describing the N-point functions characterising the large scale structure of the Universe \citep{Cooray-Sheth,asgarietal23}. In it, two point clustering statistics are expressed as a sum of two terms: 1) a one-halo term that describes pairs of points that fall within a single halo; and 2) a two-halo term that describes pairs of points that fall within different haloes.  The conceptual simplicity of the  model has made it an indispensable tool for describing clustering statistics.  However, the model is famously inaccurate in the translinear regime, exhibiting up to 20\% biases~\citep{HayashiWhite2008}.  Consequently, its utility as a tool of precision cosmology is limited~\citep{asgarietal23}.

In the standard halo model, the one halo term is typically described using an NFW \citep{NFW} or Einasto \citep{Einasto} profile, whereas the two halo term is described as the product of a halo bias times the linear matter correlation function $\xilin(r)$.  To improve the fit of the model in the translinear regime, \citet{vandenboshetal13} developed accurate fiting functions for a scale-dependent bias.  While accurate, these fitting functions provide little insight into the physics that lead to said scale dependencies.

Fortunately, our understanding of halo structure has progressed significantly over the past decade.  The first step towards our current understanding was made by \citet{diemerkravtsov14}, who demonstrated the existence of a ``splashback'' feature in the density profile of dark matter haloes.  This feature is associated with the apocentric radius of particles in their first halo orbit, and can be used as a well understood and physically-motivated halo boundary \citep{adhikari2014,moreetal2016,shellfish}.  Interestingly, if one uses the splashback radius to define the halo boundary, the resulting halo mass function is more universal than that derived using traditional halo definitions \citep{Diemer2021}.

Although physically motivated, the fact that the apocenters of splashback particles are distributed over a range of radii implies that the definition of the splashback boundary is not unique \citep{diemer2017}. \citet{GarciaRozo2021} demonstrated this feature is imprinted in the halo--mass correlation function, which can be used to uniquely define a radius--mass relation for dark matter haloes. They further found that doing so enabled them to construct a percent-level accurate --- though somewhat ad hoc --- description of the halo--mass correlation function.

Independently, \citet{fong_21} proposed an alternative halo boundary definition called the depletion radius. It is nominally defined in terms of the maximum of the mass-flow rate $I_{m}(r)=4\pi r^2\rho(r)v_r(r)$ for the inner depletion radius \citep[see also][]{cuesta_08}.  However, the infall velocity profile $v_r(r)$ is never negative at low masses. The authors note that at high mass the halo depletion radius coincides with the minimum of the ratio $\xihm(r)/\ximm(r)$, and adopt this as their definition of the inner depletion radius for low mass haloes.  The inner depletion radius roughly coincides with the splashback boundary, and was used in \citet{zhouhan23} to construct an accurate model of the halo bias $b(r,M)\equiv \xihm(r|M)/\ximm(r)$ on small and intermediate scales ($r\lesssim 20\ \hMpc$).

Building on the insights from these works, the last two years have seen a new paradigm for describing the structure of dark matter haloes begin to emerge.  \cite{diemer2022a} proposed that particles near a halo can be separated into orbiting or infalling particles according to whether or not they had experienced their first pericentric passage.  In a follow-up paper, \cite{diemer2022b} used this decomposition to arrive at fitting functions that can accurately ($\approx 5\%$) describe the density profile of dark matter haloes.

The proposal by \cite{diemer2022a} was significantly strengthened by \cite{GarciaSalazar2022}. It has long been known that the phase space distribution of particles around dark matter haloes is inherently bimodal \citep{diemand_08}.  \cite{GarciaSalazar2022} demonstrated that: 1) the two components of this phase space distribution could be cleanly separated using the accretion history of the particles; and 2) that the resulting populations of dark matter correspond to the orbiting/infall split proposed by Diemer.  They further proposed that dark matter haloes ought to be defined as the collection of all particles orbiting their self-generated potential. We refer to haloes defined in this way as \it dynamical haloes\rm.\footnote{\citet{GarciaSalazar2022} referred to these haloes as physical haloes, but we believe the ``dynamical haloes'' description is more apt.}  In this proposal \it there is no halo boundary; \rm the orbiting and infalling populations of particles overlap in phase space. Moreover: 1) the halo mass function of dynamical haloes is Press--Schechter with a slowly moving barrier; and 2) the halo bias of dynamical haloes can be accurately predicted using the peak--background split formalism \citep{desjacquesetal18}.  

The proposal by \citet{GarciaSalazar2022} has important consequences for the halo model. Specifically, because dynamical haloes are defined as the collection of all orbiting particles, it follows that the one-halo term describes the distribution of \it orbiting \rm particles in a halo, while the two-halo term describes the distribution of \it infalling \rm particles. Moreover, because the orbiting and infalling particle distributions overlap in phase space, the one- and two-halo terms necessarily overlap in configuration space.

In this work, we construct an updated halo model based on the definition of dynamical haloes proposed in \citet{GarciaSalazar2022}, focusing specifically on  the halo--mass correlation function.  We construct an accurate fitting formula for the orbiting and infall components, and arrive at a model of the halo--mass correlation function that is accurate to $\approx 2\%$ (at $1\sigma$) on scales $r\in [0.1,50]\ \hMpc$.  As detailed below, our results provide new insights into the structure of dark matter haloes and their environments.  

This paper is organised as follows.  Section~\ref{sec:data} presents the simulation we use and how the halo catalogue is constructed.  In sections~\ref{sec:orbiting_profile} and~\ref{sec:infall_profile} we characterise the orbiting and infall profiles of dark matter haloes respectively.  Section~\ref{sec:hmcf} presents our complete model for the halo--mass correlation function and characterises its accuracy.  We summarise our work and discuss some of the advantages and disadvantages of our model relative to alternatives in the literature in Section~\ref{sec:discussion}.


%% file: sections/data.tex
\section{Simulation Data and Halo Catalogue}
\label{sec:data}

This work relies on the Cold Dark Matter (CDM) simulation described in \citet{banerjeeetal20}.  The simulation was run using \gadget~\citep{Springel2005}, and contains $1024^3$ particles in a $1\ \hGpc$ box.  The softening length is $15\ \hkpc$. The cosmological parameters are $\Omega_{\rm m}=0.3$, $\Omega_\Lambda=0.70$, $n_s=0.96$, $h=0.7$, and $A_s=2\times 10^{-9}$, corresponding to $\sigma_8=0.85$.  The initial conditions are set at $z=99$ using \texttt{N-GenIC} \citep{ngenic}.  In this paper, we focus exclusively on describing the halo--mass correlation function at $z=0$.  An initial halo catalogue is obtained using the \rockstar\ halo finding algorithm \citep{Behroozi2013}.  For this initial catalogue we adopt a standard spherical overdensity definition with an overdensity threshold of 200 with respect to mean. Starting from the most massive \rockstar\ halo in the catalogue, we classify every dark matter particle within $5\ \hMpc$ of each halo as either orbiting or infalling.  The halo is then redefined as the collection of all orbiting particles, and these particles are not allowed to be members of any other halo.  We then move on to the next largest halo, and iterate as needed.  The end result is a halo catalogue where every halo is comprised exclusively of orbiting particles. We refer to these haloes as \it dynamical haloes. \rm  Additional details can be found in \citet{GarciaSalazar2022}.

\begin{figure*}
    \includegraphics[width=0.8\linewidth]{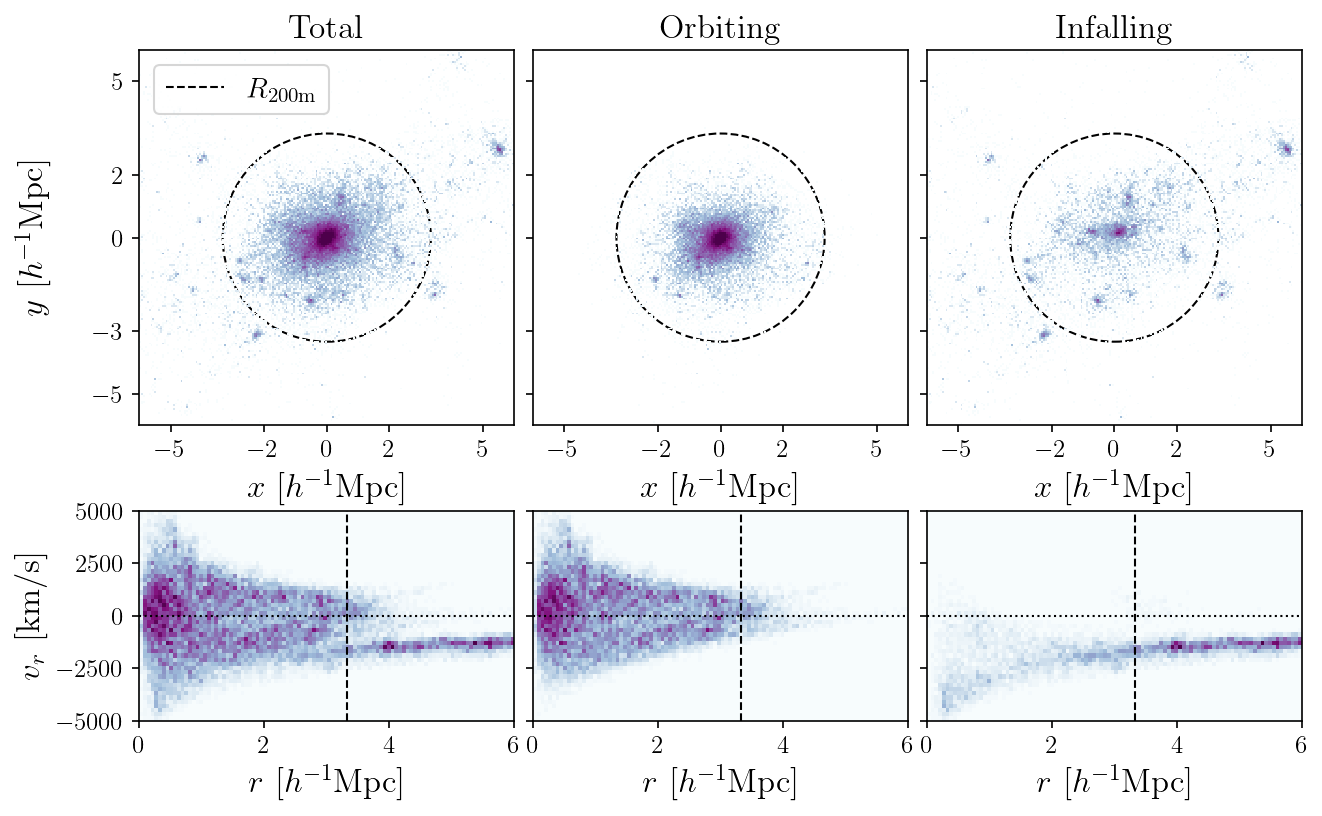}
    \caption{Dark matter particle distribution in the vicinity of the most massive halo in the simulation (left), separated into orbiting (middle) and infalling (right) populations. {\bf Top:} Configuration space $y-x$ projection. The circular dashed line denotes a standard halo radius definition $R_{200m}$ for this halo. {\bf Bottom:} Phase space $v_r-r$ projection. The vertical dashed line corresponds to $R_{200m}$.}
    \label{fig:orbinf_split_example}
\end{figure*}

Figure~\ref{fig:orbinf_split_example} illustrates how dynamical haloes differ from traditional haloes \citep[see also Fig. 3 in ][]{diemer2022a}.  The figure shows the mass distribution and phase space density of the largest halo in the simulation, with an orbiting mass $\Morb=2.35\times10^{15}\ \hmsun$. The top row shows the distribution of dark matter particles around the halo, and the bottom row shows the corresponding phase space distribution. The left column shows all the particles within a $6\ \hMpc$ box around the halo; the middle column, the orbiting particles only; and the right column, the infalling particles. We see that: 1) the phase space distribution of the halo has two distinct but overlapping components; and 2) the traditional $R_{200}$ halo boundary does not properly separate these two components. By contrast, the orbiting/infall classification proposed by \citet{GarciaSalazar2022} adequately captures the dynamical structure of the halo.

\begin{figure}
    \begin{center}
        \includegraphics[width=0.9\linewidth]{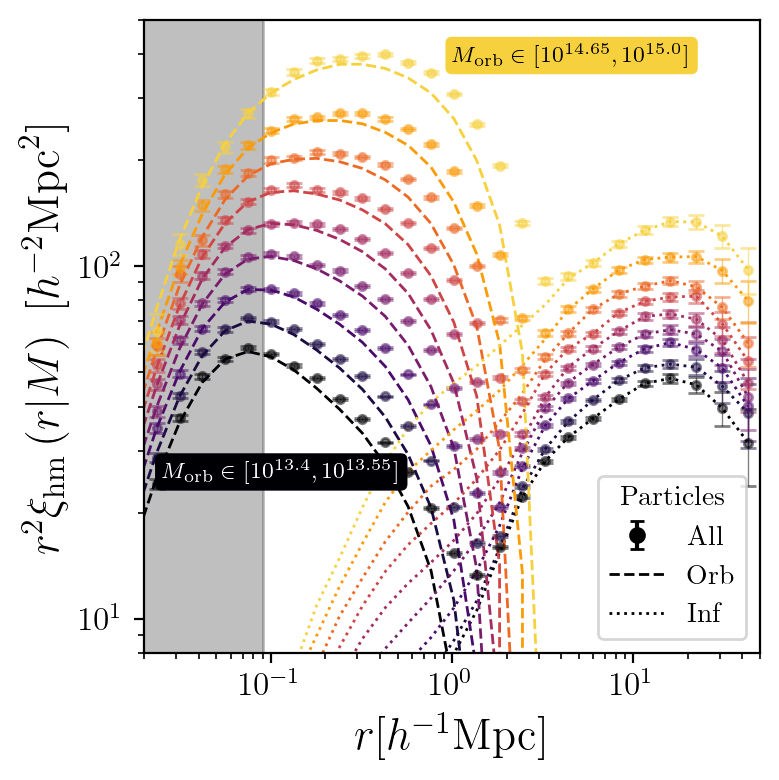}
    \end{center}
    \caption{The halo--mass correlation function of dynamical haloes as measured in our simulation. The grey region corresponds to scales smaller than six times the force softening, and are unresolved.  The dashed and dotted lines are the orbiting and infalling profiles respectively as measured in the simulations based on the orbiting/infall split proposed by \citet{GarciaSalazar2022}.  By construction, these two contributions exactly add to the full halo--mass correlation function. Errors come from jackknifing the box into 64 regions.}
    \label{fig:xihm_profile}
\end{figure}

We measure the halo--mass correlation function $\xihm(r)$ of dynamical haloes binned by orbiting mass.  When mass is measured in units of $\hMsun$, the bin edges are 
$\log_{10} M \in [$13.40, 13.55, 13.70, 13.85, 14.00, 14.15, 14.30, 14.45, 14.65, 15.00$]$. The minimum halo mass is set by the requirement that haloes contain at least 300 orbiting particles.  For reference, this is roughly equivalent to $M_{\rm 200m}\approx 2\times 10^{13}\ \hMsun$ haloes. Figure \ref{fig:xihm_profile} shows $\xihm(r)$ as measured in the simulation. We also show the orbiting (dashed line) and infalling components (dotted line) computed using the orbiting/infall split of \citet{GarciaSalazar2022}. Both components add up to $\xihm(r)$.  This decomposition is clearly evocative of the halo model, and is the primary motivation for this work.  In what follows, we describe both the orbiting and infall contributions to the halo--mass correlation function in terms of simple and, where possible, physically motivated fitting functions. 


%% file: sections/orbiting.tex
\section{The Orbiting Profile}
\label{sec:orbiting_profile}

\begin{figure*}
    \includegraphics[width=1.0\linewidth]{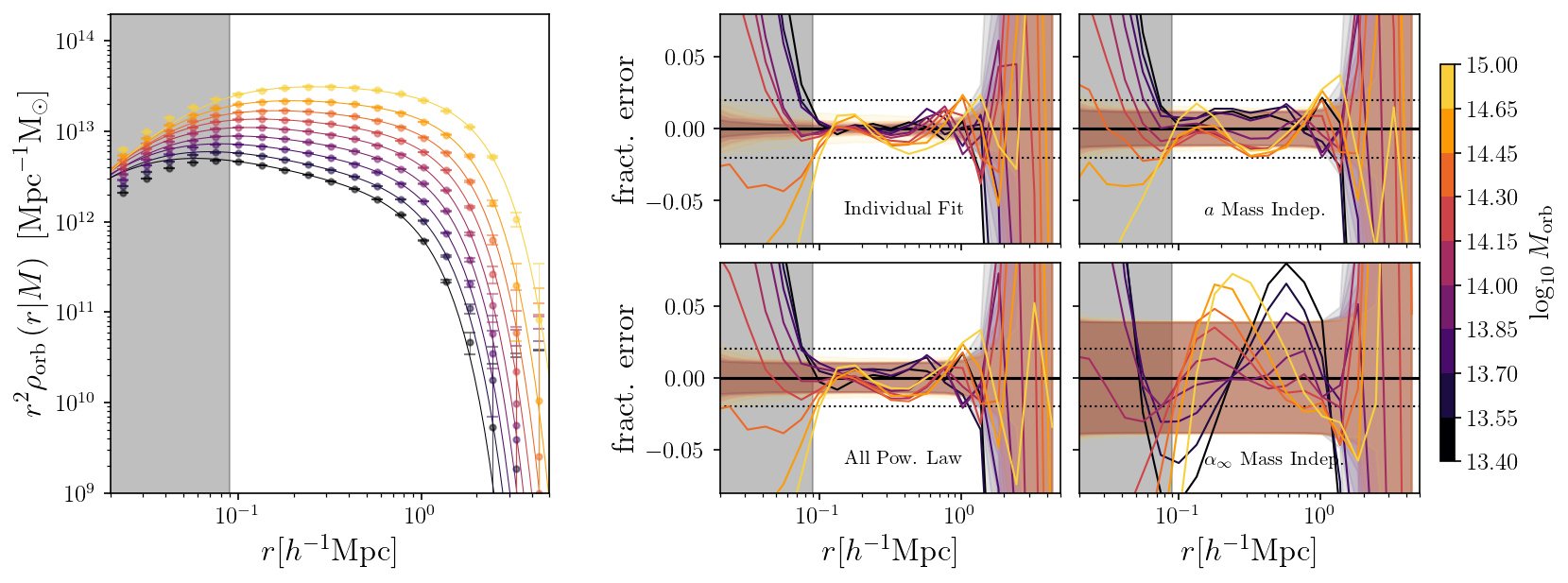}
    \caption{{\bf Left:} Orbiting density profile for stacked haloes in bins of orbiting mass. The errors bars are jackknife. The best fits for each profile are shown as solid lines. {\bf Right:} Fractional error between data and model: (top-left) each mass bin is fit individually; (bottom-left) $\rh$, $\alpha_{\infty}$ and $a$ are power laws of mass; (top-right) $\rh$ and $\alpha_{\infty}$ are power laws of mass, and $a$ is mass independent; (bottom-right) $\rh$ and $a$ are power laws of mass, and  $\alpha_{\infty}$ is mass independent. In all cases the fit is performed is restricted to scales $r\geq 90\ \hkpc$.}
    \label{fig:orb_fit}
\end{figure*}

The left panel in Figure~\ref{fig:orb_fit} shows the mean orbiting matter density profile for dynamical haloes in bins of $\Morb$. The covariance matrix for the data is obtained by jackknifing the box, and the error bars correspond to the square-root of the diagonal elements of the covariance matrix. Note that this error does \it not \rm account for any uncertainty associated with particle classification.  Consequently, we expect our error bars to be somewhat underestimated.  The grey region corresponds to a scale cut of $90\ \hkpc$, or six times the force resolution of the simulation \citep{mansfield_avestruz_20}. We see that the profiles exhibit a slowly steepening slope at small radii, and a sharp exponential decay at large radii. We can achieve an accurate fit to the data with the functional form
\begin{equation}
    \label{eq:orb_model}
    \rho_{\rm orb} (r|\Morb) = A \left(\frac{x}{a} \right)^{-\alpha(x)}\exp\left[-\frac{x^2}{2}\right]
\end{equation}
where we have defined the dimensionless radius $x=r/\rh$, and $\rh$ is the characteristic length scale associated with the exponential truncation of the orbiting profile.  The ``slope'' $\alpha(r)$ of the profile varies slowly with $r$.  We choose to parameterise this dependence so that $\alpha(r)$ extrapolates to constant values as $r\rightarrow 0$ and $r\rightarrow \infty$.  We write
\begin{equation}
    \alpha(x) = \alpha_\infty \frac{x}{a + x}
    \label{eq:slope}
\end{equation}
This functional form has $\alpha$ smoothly transitioning from $\alpha=0$ to $\alpha=\alpha_\infty$ at a scale $r=a \rh$. The value $\alpha=0$ at $r=0$ is arbitrary since these scales are not resolved in the simulation.  We chose this convention since it keeps the profile finite at $r=0$ while providing an accurate fit to the data over the scales we resolve.

There are four parameters in our model, $\{A,\rh,a,\alpha_\infty\}$.  However, the normalisation $A$ is not independent: it is set by the condition that the integral of the orbiting profile $\rhoorb(r|\Morb)$ be equal to $\Morb$. Incorporating this constraint requires some care due to the impact of resolution effects.  We describe the necessary corrections in Appendix \ref{app:mass_correction}. 

The parameter $\rh$ governs the spatial extent of the halo, so we refer to it as the \it halo radius. \rm  Note that it is the only physical scale associated with this halo profile. In particular, while the running of the slope introduces a new scale $a\rh$, we find that the parameter $a$ can be modelled as mass independent, thereby establishing $\rh$ as the sole characteristic length scale in our profile.  Finally, the parameter $\alpha_\infty$ is the asymptotic slope of the power-law component of the halo profile.  We find $\alpha_\infty$ is close to isothermal (i.e. $-2$), though it slowly decreases with increasing mass.  Consequently, our halo profiles are \it not \rm self-similar: small haloes are not just scaled versions of larger haloes.

Our choice of parameterisation appears to violate the basic finding of \citet{diemer2022b}, who argued the orbiting profile of haloes depends on two separate length scales, a scale radius $r_{\rm s}$, and a truncation radius $r_{\rm t}$.  However, the underlying phenomenology is the same. The fact that $r_{\rm s}/r_{\rm t}$ is not constant reflects the lack of self-similarity in the orbiting profile of haloes.  Indeed, the scale radius $r_{\rm s}$ is defined in terms of the logarithmic slope of the density profile, so a change in $\alpha_\infty$ in our model corresponds to a change in $r_{\rm s}$ in the \citet{diemer2022b} model.  The physical degrees of freedom are the same in both cases, but they are parameterized differently.

The likelihood of the orbiting density profile for haloes in the $k^{\rm th}$ mass bin is
\begin{equation}
    \label{eq:loglike_k}
    \ln\mathcal{L}_{k}(\rhoorb(r|\Morb)|\mathbf{\theta}) \propto -\frac{1}{2}\mathbf{D}^{\top}_{k}\mathbf{\tilde{C}}^{-1}_{\rhoorb}\mathbf{D}_{k} - \frac{1}{2}{\rm det}(\mathbf{\tilde{C}}_{\rhoorb})
\end{equation}
where $\mathbf{\theta} = \{\rh,a,\alpha_\infty\}$ is the vector of model parameters, $\mathbf{D}_{k}=\rhoorb^{data}-\rhoorb^{model}$, and 
\begin{equation}
    \label{eq:orb_covariance}
    \mathbf{\tilde{C}}_{\rhoorb} = \mathbf{C}_{\rhoorb} + {\rm diag}(\delta^{2}_{k}\rhoorb^{2})
\end{equation}
The last term in the above covariance matrix characterises the systematic uncertainty in the model.  That is, $\delta_{k}$ is the percent error we must add to account for the observed differences between our model and the data. 

The best fit model to the orbiting profile for each individual halo mass bin is shown as solid lines in the left panel of Figure~\ref{fig:orb_fit}.  The residuals from the fit are shown in the top-left residuals sub-panel.  The horizontal dotted lines in the residual panels correspond to a 2\% difference between the data and the best fit model. 

\begin{figure*}
    \includegraphics[width=0.9\linewidth]{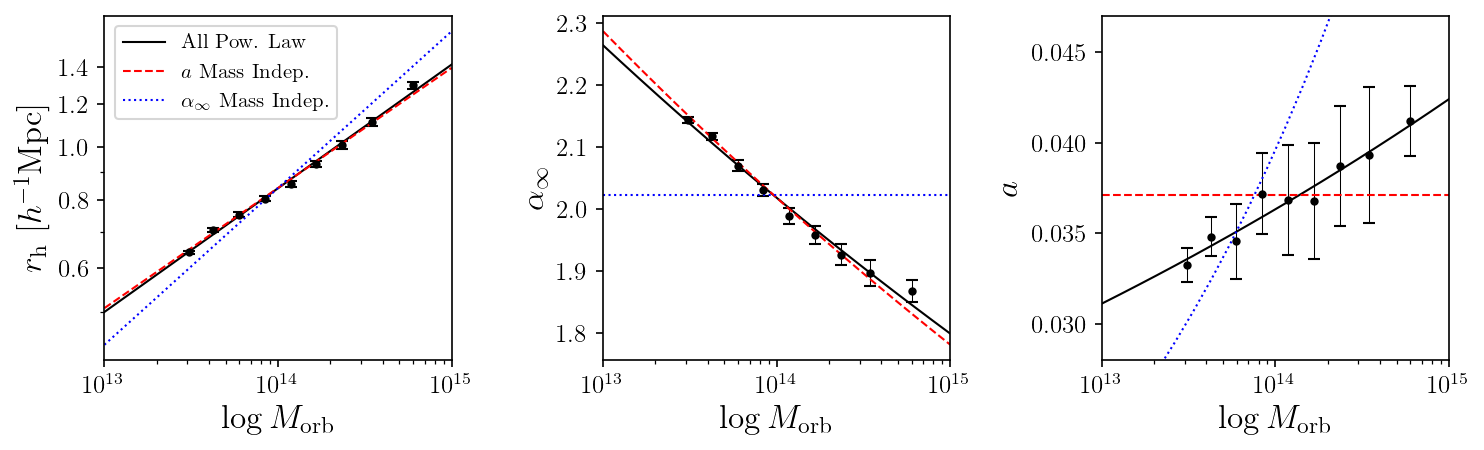}
    \caption{Best-fit orbiting profile model parameters as a function of halo mass. The data points correspond to the individual profile fit maximum likelihood values, and the error bars are the 68\% confidence interval. The solid lines are not a fit to these data points, but rather the best fit smooth model to the orbiting profile for three smooth parameterisations, as labelled.}
    \label{fig:orb_fit_pars}
\end{figure*}

Figure~\ref{fig:orb_fit_pars} shows how each of our model parameters scales with mass. Unsurprisingly, $\rh$ increases with increasing mass: the more massive a halo, the larger it is.  Note that while the shape parameter $a$ controls the detailed shape of the halo profile at small scales, the scale $r=a\rh$ is unresolved in our simulations, being roughly three times the softening length. Consequently, we do not ascribe it physical significance; it is simply a means of parameterising the running of the power-law index of the halo profile over the scales we resolve.

To construct a model for the density profile as a function of mass and radius, we parameterise the halo--mass dependence of each of our model parameters as a power law in mass.  We then simultaneously fit all mass bins for the amplitudes and slopes of these power-law relations. We will refer to this parameterisation as the \textit{smooth model}. The power laws for our model parameters are given by
\begin{align}
    \rh(M_{\rm orb}) & = r_{{\rm h}, p} \left(\frac{M_{\rm orb}}{M_{\rm p}}\right)^{r_{{\rm h}, s}}
\end{align}
\begin{align}
    \alpha_\infty(M_{\rm orb}) & = \alpha_{\infty, p} \left(\frac{M_{\rm orb}}{M_{\rm p}}\right)^{\alpha_{\infty, s}}, \\
    a(\Morb) & = a_{p} \left(\frac{M_{\rm orb}}{M_{\rm p}}\right)^{a_{s}}.
\end{align}
The likelihood of the smooth orbiting density profile is then
\begin{equation}
    \label{eq:loglike}
    \ln\mathcal{L}(\{\rhoorb(r|\Morb)\}_{k=1}^{N}|\mathbf{\theta}) = \sum_{k}\ln\mathcal{L}_{k}
\end{equation}
where $\mathbf{\theta} = \{r_{{\rm h}, p}, r_{{\rm h}, s}, \alpha_{\infty, p}, \alpha_{\infty, s}, a_{p}, a_{s}, \delta\}$ is the vector of model parameters. We use a single parameter $\delta$ to characterise any required excess variance across all mass bins. We set a flat prior $(-\infty, 0]$ for $\alpha_{\infty, s}$, and $[0, \infty)$ on the rest. Moreover, we assume no correlation between mass bins. We chose a pivot mass $M_{\rm p}=10^{14}\ \hmsun$, which roughly decorrelates the amplitude and slope parameters in our fits. 

The best fit smooth model obtained when fitting all bins simultaneously is nearly indistinguishable from the individual best fits, and is therefore not shown in Figure~\ref{fig:orb_fit}. The corresponding residuals are shown in the bottom-left sub-panel.  The band about zero characterises the error in our fits, and includes the best fit value for the added statistical uncertainty parameter $\delta$. Evidently, the smooth model remains accurate to better than 2\% across all masses. Indeed, the best fit value for $\delta$, which characterises the accuracy of our model, is $\delta=1.0\%$. The smooth model predictions for each of the orbiting profile parameters $\rh$, $\alpha_\infty$, and $a$ are shown as black solid lines in Figure~\ref{fig:orb_fit}.  Note these lines are not fits to the data points in those panels, but rather fits to the complete set of stacked orbiting halo profiles.  

Traditional halos are typically defined using an overdensity criterion, which leads to a radius--mass relation with a slope of 1/3.  By contrast, we find that the slope of the $\rh$--$\Morb$ relation is shallower than $1/3$ at $3\sigma$: $r_{\rm h},s=0.226\pm 0.003$.  

Turning to the power laws for $\alpha_{\infty}$ and $a$ as a function of mass are both very close to zero. For this reason, we also test the accuracy of our model for two other cases: 1) $\alpha_{\infty}$ is a power law in mass and $a$ is mass independent; and 2) $a$ is a power law in mass and $\alpha_{\infty}$ is mass independent. The residuals from the best fit model in each of these two cases are shown in the top-right and bottom-right sub-panels of Figure~\ref{fig:orb_fit} respectively.  The corresponding curves for these parameters as a function of mass are also shown in Figure~\ref{fig:orb_fit_pars}.

From the right-most residual panels in Figure~\ref{fig:orb_fit} it is clear that making $\alpha_\infty$ mass independent severely degrades the quality of the fits.  Conversely, modelling the parameter $a$ as mass independent hardly impacts the quality of our best fit model. Given that the parameter $a$ corresponds to an unresolved scale, and that making this parameter mass-dependent results in negligible improvement in the quality of our fits, we have chosen to model $a$ as mass-independent.

In summary: the orbiting profile of dynamical haloes can be accurately ($\approx 1\%$) described as an exponentially truncated power-law (equation~\ref{eq:orb_model}).  The halo radius $\rh$ characterises the exponential truncation, and is the only dimension-full parameter in our model.  As expected, $\rh$ increases with mass, but the slope of the relation is significantly shallower than the traditional value of $1/3$.  The slope of the power-law part of the density profile varies slowly with radius as per equation~\ref{eq:slope}.  This introduces 2 new parameters: $a$, and $\alpha_\infty$.  Of these, $\alpha_\infty$ scales with mass as a power-law, whereas $a$ can be modelled as mass independent. Thus, the total number of free parameters in our model is three: two parameters $\rh$ and $\alpha_\infty$ that depend on mass, and a population parameter $a$.  The maximum likelihood values for the parameters in our fiducial (constant $a$) smooth model are shown in Table~\ref{tab:parameters}.  The corresponding posterior distributions are shown as the red contours in Figure~\ref{fig:xihm_fit_corner}.  


%% file: sections/infall.tex
\section{The Infall Profile}
\label{sec:infall_profile}

\begin{figure*}
    \includegraphics[width=0.85\linewidth]{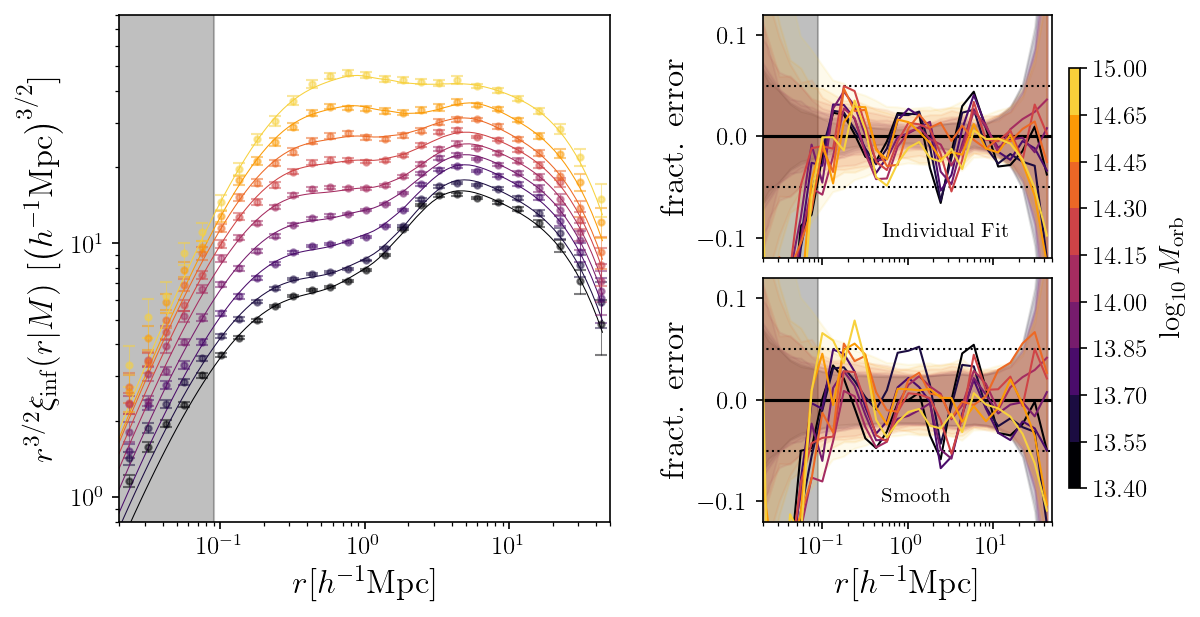}
    \caption{{\bf Left:} Infalling halo--mass correlation function for haloes in bins of orbiting mass. The errors bars are jackknife. The best fits for each profile are shown as solid lines. {\bf Right:} Fractional error between data and model when fitting each bin individually (top),  or when performing a joint fit to all mass bins using our smooth parameterisation (bottom). In all cases the fit is performed on scales above the resolution limit of the simulation ($r\geq 90\ \hkpc$, shown in grey).}
    \label{fig:inf_fit}
\end{figure*}

We wish to characterise the infall contribution to --- or, equivalently, the two-halo term of --- the halo--mass correlation function.  Since our orbiting/infall split extends only to a radius of $5\ \hMpc$ away from a halo, we assume all particles at distances larger than this are infalling. This is already an excellent approximation for all haloes at $r\approx 5\ \hMpc$.  For scales less than $5\ \hMpc$, the infall correlation function is defined as $\rho_{\rm inf}(r)/\bar \rho_{\rm m}-1$.  On scales larger than this we set $\xiinf=\xihm$.  We measure the halo--mass correlation function up to a maximum radius of $50\ \hMpc$. On larger scales, the measurements become noisy due to the cosmic variance in the simulation box. The left panel of Figure~\ref{fig:inf_fit} shows the product $r^{1.5}\xiinf$ as a function of radius for each of our halo mass bins. The vertical axis is multiplied by $r^{1.5}$ to reduce the dynamical range of the $y$-axis and accentuate the non-power-law features in the correlation function. All errors are jackknife. Our goal is to fit this data.


\subsection{The Large Scale Limit of $\xiinf$}
\label{sec:large_scale_limit}

\begin{figure}
    \begin{center}
        \includegraphics[width=0.9\linewidth]{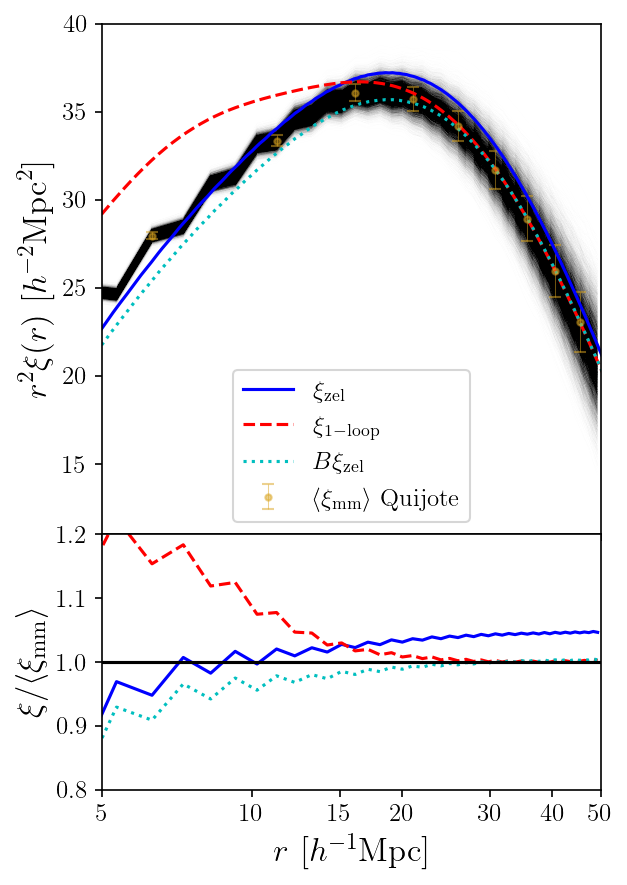}
    \end{center}
    \caption{{\bf Top}: The mass--mass correlation function measured for 15,000 realisations of Quijote's fiducial cosmology at $z=0$. The points with error bars correspond to the ensemble average, and the error bars show the cosmic variance in a single box, as opposed to the error on the mean. The lines correspond to the Zel'dovich correlation function, the correlation function $B\xizel$ where $B$ is a constant multiplicative offset chosen to match the non-linear correlation function at $r\approx 50\ \hMpc$ scales, and the 1-loop LPT prediction calculated using the code {\sc velocileptors} \citep{Chen2020}. {\bf Bottom}: Ratio of these various correlation functions to the mean correlation function measured in Quijote. 
    }
    \label{fig:ximm_quijote}
\end{figure}

Figure~\ref{fig:inf_fit} demonstrates that the infall correlation function is feature-rich.  Traditionally, one sets the two halo term of $\xihm$ to $\xihm^{\rm 2-halo}= b\xilin$, where $\xilin$ is the linear correlation function. More generally, on large scales one can set
\begin{equation}
    \xiinf = b\xilarge.
    \label{eq:xilarge0}
\end{equation}
for an appropriate model of $\xilarge$. Indeed, the ratio $\xihm/\ximm$ is very nearly constant for $r\in [10,50]\ \hMpc$ \citep{GarciaSalazar2022}. Consequently, we seek a function $\xilarge$ that provides an accurate description of the matter correlation function on the scales resolved by this work.

When defining $\xilarge$ we could rely on existing approximations for $\ximm$, for instance {\sc halofit} \citep{halofit,smith_angulo19,bartlett_etal24} or emulators \citep{coyote,dark_emulator,anguloretal21,mira_titan23}.  However, we wish to make our model as interpretable as possible.  We had anticipated using Effective Field Theory (EFT) models for $\xilarge$ \citep[e.g.][and others]{Porto_2014,Vlah_2015, Chen2020}, but found these models ill-suited for our purposes.  Specifically, figure~\ref{fig:ximm_quijote} compares the $z=0$ mass correlation measured using the 15,000 {\it Planck} simulations of the Quijote simulations suite \citep{Quijote_sims} with the best-fit EFT model based on one-loop Lagrangian Perturbation Theory (LPT).  We see that on scales $r\lesssim 20\ \hMpc$, the LPT model: 1) deviates strongly from the simulation data; and 2) has obvious features not present in the simulation data.  If we were to choose a one-loop LPT description of $\xilarge$, our small scale parameterization for $\xiinf$ would have to ``undo'' these features.

For these reasons, we rely instead on the empirical approximation $\ximm \approx B\xizel$ where $B=0.948$ and $\xizel$ is the Zeldovich correlation function.  As shown in Figure~\ref{fig:ximm_quijote}, this approximation is accurate to better than 1\% over the scales $r\in[20,50]\ \hMpc$.  Moreover, the failure of this approximation at small scales is both featureless and less severe than that of the one-loop LPT model, making it a more useful model for our purposes. Consequently, we define
\begin{equation}
    \xilarge \equiv B\xizel.
    \label{eq:large}
\end{equation}
where $B=0.948$ is held fixed. The inequality $B<1$ is due to the impact of non-linear growth on $\ximm$, as evidenced by the fact that: 1) the prefactor $B$ is a function of redshift, with $B\rightarrow 1$ as redshift increases; and 2) on scales $r\in [30,50]\ \hMpc$, the one-loop LPT predictions for $\ximm$ (with a counterterm) agrees with the approximation $B\xizel$ with better than 1\% accuracy.  We caution that on very large scales one \it must \rm recover the perturbative limit $\ximm \rightarrow \xizel$ as $r\rightarrow \infty$, or $B=1$.  Consequently, \it we anticipate that our model does not currently extrapolate to linear scales. \rm  We will address this problem in future work.

\subsection{The Small Scale Limit of $\xiinf$}
\label{sec:small_scale_limit}

On small ($r\lesssim 10\ \hMpc$) scales $\xihm$ deviates from both $\ximm$ and $\xilarge$ in a scale dependent way.  This leads us to define a scale dependent bias $\beta(r)$ such that $\xiinf = b\beta(r)\xilarge$.  The function $\beta(r)$ only has an impact on small scales, and is therefore distinct from the higher order parameters appearing in perturbative expansions (e.g. $b_2$, $b_{\rm s^2}$, $b_3$).  We find empirically that $\beta(r)$ increases with decreasing radius, which suggests $\beta$ should take the form $1+(\rinf/r)^\gamma$.  To avoid divergences at $r=0$ we adopt a cored power-law instead, and hypothesise that the core radius should scale with the halo radius $\rh$.  These assumptions lead us to write 
\begin{equation}
    \beta(r) = 1+\left[\dfrac{\rinf}{\mu\rh+r}\right]^{\gamma}.
    \label{eq:beta}
\end{equation}
Because $\rinf$ characterises the deviations from linear bias due non-linear growth, we anticipate $\rinf$ is related to the non-linear scale $\kNL$ defined via $\Delta^2_{\rm lin}(\kNL)=1$. Indeed, we find $\rinf$ is independent of halo mass, with $\kNL\rinf \approx 2/3$.

\begin{figure}
    \begin{center}
        \includegraphics[width=\linewidth]{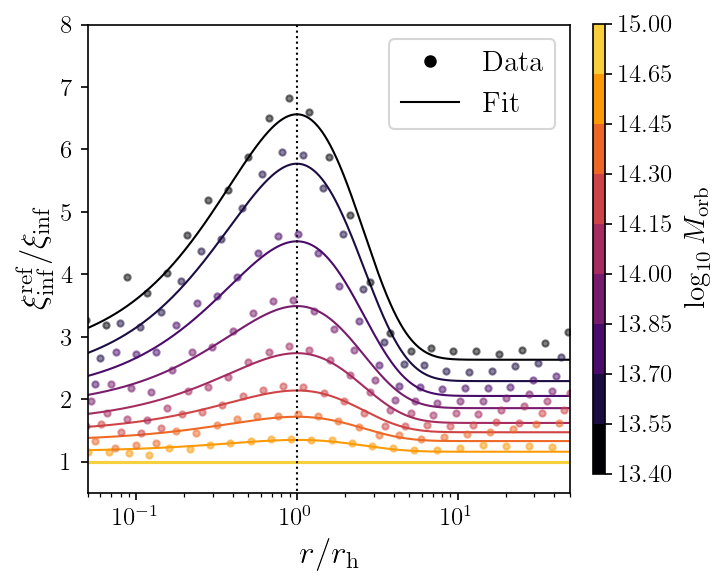}
    \end{center}
    \caption{Ratio of the infalling halo--mass correlation function of the largest mass bin to all other mass bins. A ``bump'' appears at $r=\rh$ as we decrease in mass, becoming more prominent for the lowest masses. The solid lines are a fit to the data points using Eq. \eqref{eq:ratio}.}
    \label{fig:inf_ratio}
\end{figure}

Our model is not yet complete, however.  Specifically, we find that the orbiting profile imprints itself onto the distribution of infalling matter at small scales.  This is shown in Figure~\ref{fig:inf_ratio}, where we plot the ratio $\xiinf(r|M_{\rm ref})/\xiinf(r|M)$ where $M_{\rm ref}$ is our most massive bin.  This ratio is then plotted as a function of $r/\rh$, where $\rh$ is the halo radius of haloes of mass $M$. Evidently, the halo radius $\rh$ is imprinted onto the distribution of infalling particles, further cementing $\rh$ as the characteristic radial scale for the halo.  

To fit the ratios in Figure~\ref{fig:inf_ratio} we define $x\equiv r/\rh$, and then fit the ratio using the function
\begin{equation}\label{eq:ratio}
   f(x) = C + \eta xe^{-x}
\end{equation}
The fits are shown as solid lines in Figure~\ref{fig:inf_ratio}. The vertical displacement parameter can be set to $C=1$ by fitting to the ratio $b_{\rm ref}\xiinf^{\rm ref}/b\xiinf$, where $b$ is the large scale bias parameter discussed above. This means that we only need one additional parameter ($\eta$) to fully characterise this feature in the infall correlation function.

Our final model for the infall correlation function is
\begin{equation}
    \label{eq:inf_model}
    \xiinf(r) = \frac{\beta(r)}{1+\eta xe^{-x}}b\xilarge(r).
\end{equation}
where $\beta(r)$ is given by Eq.~\ref{eq:beta}, and $x\equiv r/\rh$.  The parameter $\eta$ characterises the amplitude of the impact of the orbiting particles on the infall profile.

Our final model for $\xiinf$ has five parameters: $\{b, \eta, \rinf, \gamma, \mu\}$, each of which has a simple physical interpretation: $b$ is the large scale halo bias; $\eta$ characterises the amplitude of the ``dip'' in the infall term due to the gravitational potential generated by the orbiting mass of the halo; $\rinf$ and $\gamma$ characterise the power-law behaviour of the small scale dependent bias; and $\mu$ characterises the scale at which $\beta(r)$ reaches its maximum value. Note that in this section we do not consider $\rh$ to be a model parameter that needs to be fit for.  Rather, we adopt the best fit value for $\rh$ obtained from fitting the orbiting profiles. 

\subsection{Infall Fits}
\label{sec:infall_fits}

Our final model for $\xiinf$ is set by Eq.~\eqref{eq:inf_model} where $\xilarge$ is defined via Eq.~\eqref{eq:large}.  We fit the simulation data to our model for $\xiinf$ using the likelihood in Eq.~\eqref{eq:loglike_k}, where we simply exchange $\rhoorb$ for $\xiinf$. Because of the way our measurements were made, the covariance matrix of the extended infalling correlation function, $\mathbf{C}_{\xiinf}$, is a block-diagonal matrix, as we had no easy way to estimate the covariance between the $r \leq 5\ \hMpc$ and $r\geq 5\ \hMpc$ for $\xiinf$. We set flat priors $[0, \infty)$ on all the parameters. Our best fit model is shown as the solid lines in the left panel of Figure~\ref{fig:inf_fit}.  The top-right panel of the same figure shows the fractional residuals between the data and our model.  We see that our model is accurate to 5\% or better. 
 
\begin{figure*}
    \centering
    \includegraphics[width=\linewidth]{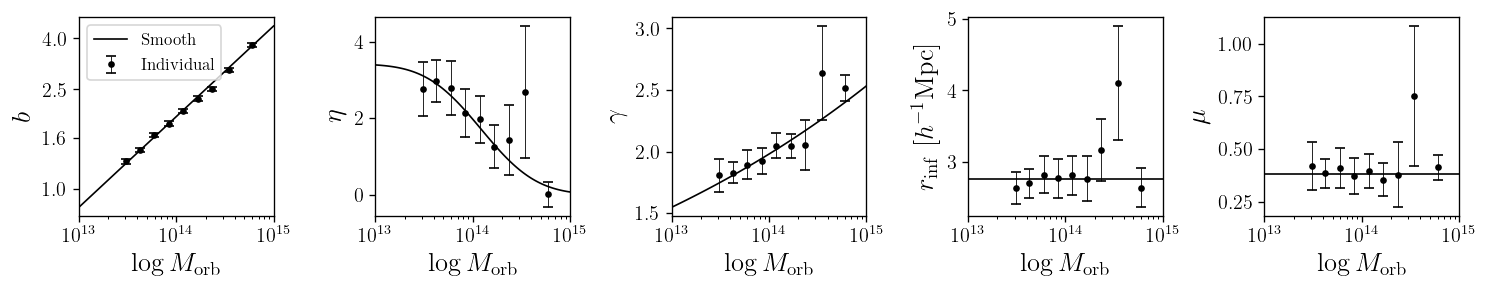}
    \caption{Infalling profile model parameters. The data points correspond to the individual profile fit maximum likelihood values, and the error bars are the 68\% confidence interval. The solid lines are not a fit to the data points shown above, but rather the best fit smooth model to the infalling profile.}
    \label{fig:inf_pars}
\end{figure*}

Figure~\ref{fig:inf_pars} shows the maximum likelihood estimates for each of our model parameters as a function of $\Morb$.  The error bars represent the 68\% confidence regions obtained from the parameter posteriors.  Remarkably, both $\rinf$ and $\mu$ are mass-independent.  The fact that $\rinf$ is mass independent supports the notion that it is related to the non-linear scale $\kNL$.  Likewise, the fact that $\mu$ is mass independent emphasises the fact that $\rh$ is the unique radial scale associated with haloes of a given orbiting mass.

We now turn to parameterise a smooth model for $\xiinf$ as a function of mass. We parameterise the mass dependence of $b(\Morb)$ and $\gamma(\Morb)$ with power laws.\footnote{Because our model does not correctly extrapolate to $r\rightarrow \infty$, we cannot identify our bias parameters $b$ with the large scale bias predicted by the peak--background split, leading us to rely on the power-law approximation for $b(M)$.}  The mass dependence of $\eta(\Morb)$ is more complicated, and is modelled using an error function.\footnote{We also tried using a power-law description for $\eta(\Morb)$, but failed to achieve good fits with that parameterisation.}  Both $\rinf$ and $\mu$ are modelled as being mass independent. Our updated model parameters for this smooth model are given by
\begin{align}
    b & = b_{p} \left(\frac{M_{\rm orb}}{M_{\rm p}}\right)^{b_{s}},  \\
    \gamma & = \gamma_{p} \left(\frac{M_{\rm orb}}{M_{\rm p}}\right)^{\gamma_{s}}, \\
    \eta & =  \frac{\eta_0}{2}\left[1-{\rm erf}\left({\frac{\log_{10}\left[\frac{M_{\rm orb}}{M_{\rm p}}\right] - \eta_{m}}{\eta_{\sigma}}}\right)\right]
\end{align}
We fit for the smooth model parameters and again choose a pivot mass $M_{\rm p}=10^{14}\ \hmsun$. The likelihood of the smooth infalling correlation function is given by Eq. \eqref{eq:loglike} with the appropriate exchange $\rhoorb\rightarrow\xiinf$.  Likewise, the vector of model parameters is now $\mathbf{\theta} = \{b_{p}, b_{s}, \gamma_{p}, \gamma_{s}, \eta_{0}, \eta_{m}, \eta_{\sigma}, \rinf, \mu\}$, i.e. we describe our 9 $\xiinf$ curves with one parameter per curve. We assume no correlation between mass bins when performing our simultaneous fit. 
The maximum likelihood parameter values for our smooth $\xiinf$ model relations are shown in Table~\ref{tab:parameters}. Our smooth mass model is nearly indistinguishable from the single-bin fits in the left panel of Figure~\ref{fig:inf_fit}, so we do not show them there. The percent residuals for our best fit smooth model are shown in the right-most panel of Figure~\ref{fig:inf_fit}.  The corresponding curves for the single-bin parameters as a function of mass are shown as solid lines in Figure~\ref{fig:inf_pars}. The posterior distribution of our infall parameters is shown as blue contours in Figure~\ref{fig:xihm_fit_corner}. The accuracy of our best fit model is characterised using the excess variance parameter $\delta$, for which we recover $\delta=2.1\%$.  The single most discrepant point in our model is $\approx 5\%$ offset from our model fit.

In summary: our infall model assumes $\xiinf \propto b\ximm$, where $\ximm$ is described using the empirical approximation $\ximm\approx B\xizel$ on scales $r\in [10,50]\ \hMpc$.  To accurately describe smaller scales we introduce a scale-dependent bias $\beta(r)$ which we model as cored power-law.  This introduces three new parameters: the amplitude and slope of the power-law ($\rinf$ and $\gamma$), and the core radius. The core radius is a constant multiple $\mu$ of the halo radius.  Finally, we found the orbiting density profile imprints itself in the infall profile, which necessitates introducing one additional free parameter ($\eta$) that characterises the amplitude of this effect.  The total number of parameters required to describe $\xihm$ is three mass-dependent parameters ($b$, $\gamma$, and $\eta$), plus two population parameters $\mu$ and $\rinf$ that apply to all halos.  Our model describes the infall profile on all scales with $\approx 2\%$ accuracy.


%% file: sections/halo_matter.tex
\section{The halo--mass Correlation Function}
\label{sec:hmcf}

\subsection{Results}

Following the halo definition advocated by \citet{GarciaSalazar2022}, one can identify the one- and two- halo terms of the halo--mass correlation function as those arising from the orbiting and infalling population respectively.  Note that because all particles in a simulation are classified, this decomposition inherently occurs in density space.  That is, we ought to write $\rho_{\rm tot}=\rhoorb+\rhoinf$. The correlation function takes the form $\xihm(r)=\rho_{\rm tot}/\bar \rho_{\rm m}-1$, so there is some ambiguity as to whether the ``-1'' goes with the orbiting term, the infall term, or both.  In configuration space, it is most natural to assign the ``-1'' to the infall term, as we have done so far.  With this convention, our model for the halo--mass correlation function can be written as
\begin{equation}
    \xihm(r) = \frac{\rhoorb(r)}{\barrhom} + \xiinf(r).
    \label{eq:full_model}
\end{equation}

\begin{figure*}
    \centering
    \includegraphics[width=0.80\linewidth]{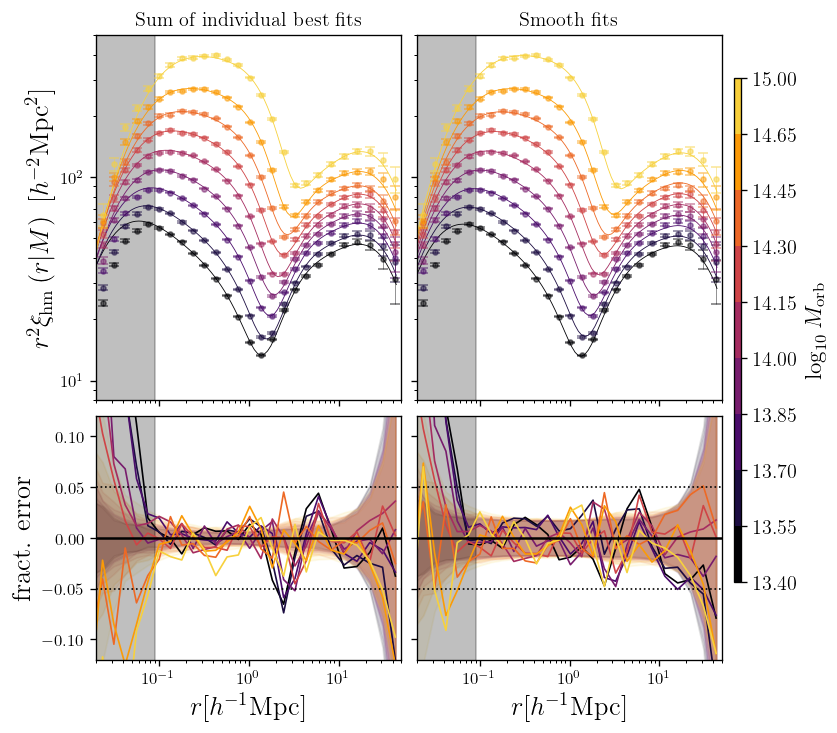}
    \caption{Halo--mass correlation function in bins of orbiting mass. The errors bars are jackknife. {\bf Left:} The solid lines are the sum of the best fit orbiting and infalling profiles, and are not a fit to the points in this figure. {\bf Right:} Our best fit smooth model to the data points in the Figure.  {\bf Bottom:} Fractional error between data and model. In both cases the fit is performed above the shaded region corresponding to $90\ \hkpc$.  The coloured band on the residuals plot on the left shows the jackknife error bars of the data, while the plot on right adds in quadrature the best-fit scatter parameter $\delta$ that characterises the accuracy of the fit.}
    \label{fig:xihm_fit}
\end{figure*}

The left panel of Figure~\ref{fig:xihm_fit} shows the halo--mass correlation function for each of our mass bins as data points with error bars.  All errors come from jackknifing the box.  The solid lines correspond to the sum of the orbiting and infall terms predicted from out best bin-by-bin fits of the orbiting and infall profiles discussed in sections~\ref{sec:orbiting_profile} and \ref{sec:infall_profile}.  The fractional errors from this model are shown in the bottom-left panel of the same figure, and are typically $\lesssim 2\%$, though the worst fit points exhibit residuals as large as $\approx 8\%$.  

Critically, however, the model curves in the left panel of Figure~\ref{fig:xihm_fit} are \it not \rm a fit to these data.  These are the model curves defined by the fits to the orbiting and infalling components from sections~\ref{sec:orbiting_profile} and \ref{sec:infall_profile}, which in turn rely on the orbiting/infall split adopted in \citet{GarciaSalazar2022}. It is therefore possible that errors in particle classification compromise the accuracy of our model.

To test for this possibility, we use our full model (orbiting+infalling) to fit the halo--mass correlation function directly.  Note that while our functional form for $\xihm$ model was informed by the orbiting and infalling profiles, the fit is completely agnostic to how or even if the particles are classified.  The resulting model fit is shown in the right hand panel of Figure~\ref{fig:xihm_fit}, and the corresponding residuals are shown in the bottom-middle column. The accuracy of our model is characterised be the excess variance needed to fit the data, for which we find $\delta=1.62\%$. The worst-fit data point differs from the model by $\approx 5\%$.    

\begin{table*}
   \centering
   \caption{Maximum likelihood estimates of the smooth model parameters for the orbiting, infalling and $\xihm$ profiles. The confidence intervals ($\approx\,68\%$) have been set to the largest absolute value of the two bounds.}
   \begin{tabular}{r|l|c|c|c}\hline
        Parameter & Description & $\rhoorb$ & $\xiinf$ & $\xihm$\\\hline
        $r_{{\rm h}, p}$ & Halo radius pivot ($\hkpc$) & $840.3 \pm 2.2$ & --- & $818.6 \pm 4.8$ \\
        $r_{{\rm h}, s}$ & Halo radius power & $0.226   \pm 0.003$ & --- & $0.244 \pm 0.005$ \\
        $\alpha_{\infty, p}$ & Asymptotic slope pivot & $2.018 \pm 0.002$ & --- & $1.991 \pm 0.006$ \\
        $\alpha_{\infty, s}$ & Asymptotic slope power & $-0.050 \pm 0.001$ & --- & $-0.048 \pm 0.002$ \\
        $a$ & Inner scaling & $0.037 \pm 0.001$ & --- & $0.036 \pm 0.001$ \\ \hline
        $b_{p}$ & Large-scale bias pivot & --- & $1.948 \pm 0.012$ & $1.888 \pm 0.014$ \\
        $b_{s}$ & Large-scale bias power & --- & $0.362 \pm 0.004$ & $0.375 \pm 0.005$ \\
        $\gamma_{p}$ & Power law slope pivot & --- & $1.979 \pm 0.028$ & $1.526 \pm 0.0.61$ \\
        $\gamma_{s}$ & Power law slope power & --- & $0.107 \pm 0.003$ & $0.168 \pm 0.014$ \\
        $\eta_{0}$ & Dip parameter amplitude & --- & $3.422 \pm 0.266$ & $1.265 \pm 0.207$ \\
        $\eta_{m}$ & Dip parameter mean & --- & $0.088 \pm 0.024$ & $0.223 \pm 0.055$ \\
        $\eta_{\sigma}$ & Dip parameter spread & --- & $0.634 \pm 0.046$ & $0.456 \pm 0.078$ \\
        $r_{\rm inf}$ & Non-linear scale ($\hMpc$) & --- & $2.760 \pm 0.066$ & $2.261 \pm 0.081$ \\ 
        $\mu$ & Power law core & --- & $0.381 \pm 0.020$ & $0.094 \pm 0.049$ \\ \hline
        $\log_{10}\delta$ & \% error in quadrature & $-2.010 \pm 0.033$ & $-1.665 \pm 0.022$ & $-1.790 \pm 0.020$ \\\hline
   \end{tabular}
   \label{tab:parameters}
\end{table*}

\subsection{The Impact of Particle Classification}

\begin{figure*}
    \includegraphics[width=0.95\linewidth]{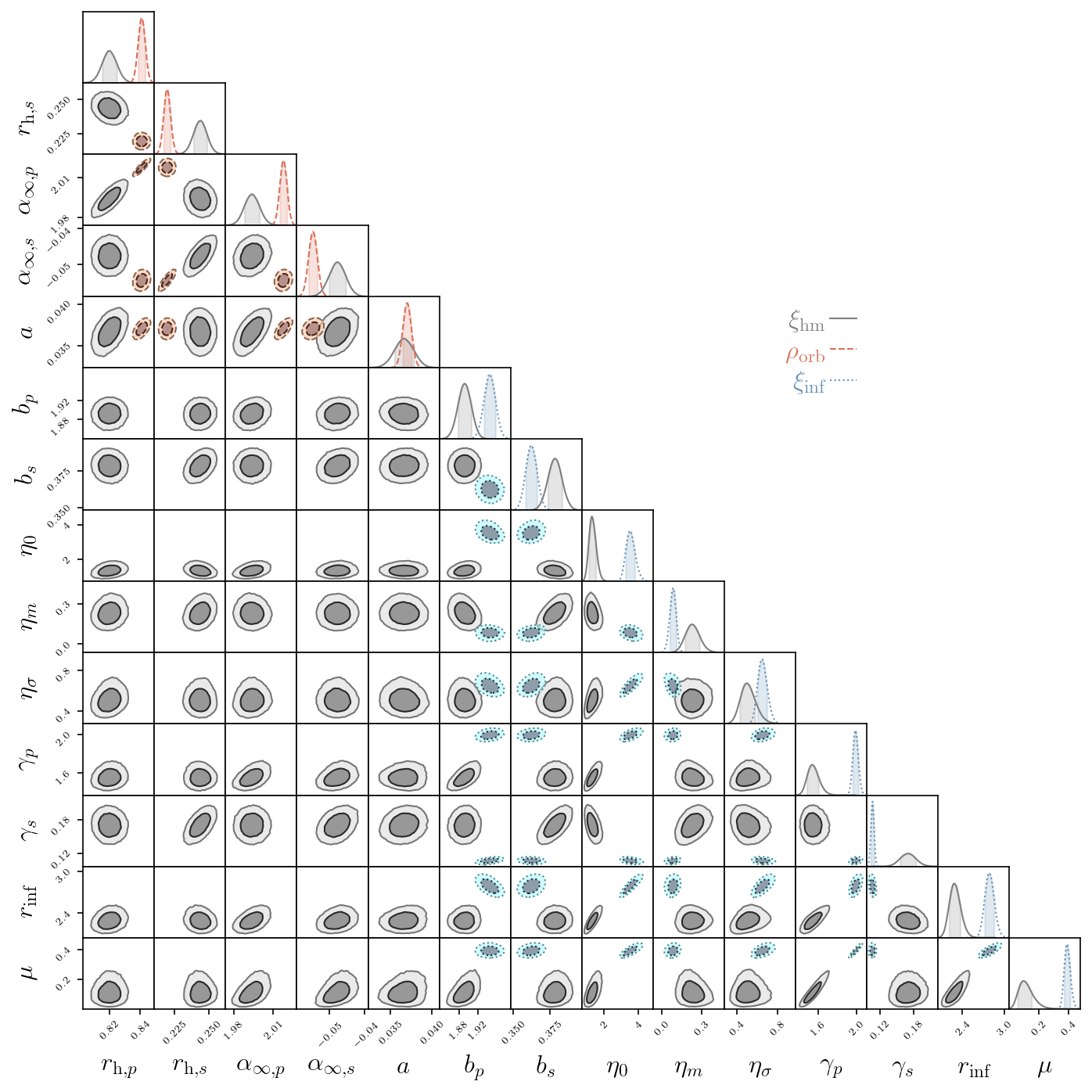}
    \caption{68\% and 95\% posterior distributions for the parameters our smooth model for the halo--mass correlation function $\xihm$. The grey-solid contours come from fitting $\xihm$, whereas the red-dashed and blue-dotted contours show the fits to the orbiting and infall components when particles are separated using the criterion of \citet{GarciaSalazar2022}.}
    \label{fig:xihm_fit_corner}
\end{figure*}

Figure~\ref{fig:xihm_fit_corner} compares the smooth model parameter posteriors obtained using two different analysis: 1) fitting $\xihm$ directly; and 2) fitting $\rhoorb$ and $\xiinf$ independently as per sections~\ref{sec:orbiting_profile} and \ref{sec:infall_profile}.  Evidently, the posterior distributions do \it not \rm agree \citep[see also][]{diemer2022b}. These results suggest the presence of systematic errors in the orbiting/infall decomposition of \citet{GarciaSalazar2022}.

\begin{figure*}
    \includegraphics[width=\linewidth]{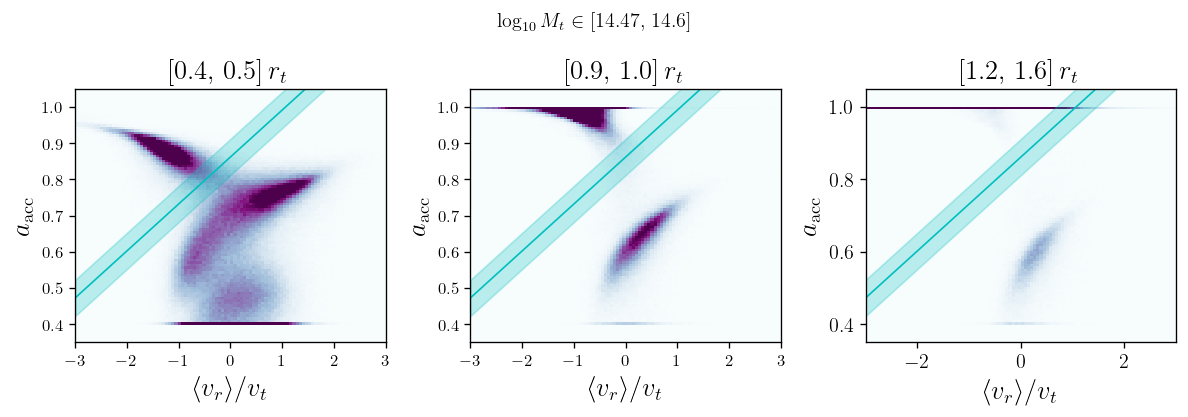}
    \caption{Criteria for splitting particles into orbiting and infalling.  The solid cyan line generates the fiducial split outlined in \citet{GarciaSalazar2022}. The band is at $\pm0.05$ from the fiducial line. For definitions of the truncation mass $M_t$ and radius $r_t$ see \citet{GarciaRozo2021, GarciaSalazar2022}. We adopt the width of these bands as a rough estimate of the systematic uncertainty associated with orbiting/infall split methodology of \citet{GarciaSalazar2022}.}
    \label{fig:split_accvr}
\end{figure*}

To further test this hypothesis we have evaluated the impact of small, plausible modifications to the \citet{GarciaSalazar2022} algorithm for particle classification. 
In that work, particles are classified by applying a cut in the space of accretion time $\aacc$ vs. mean radial velocity $\avg{v_r}$ \citep[see][for details]{GarciaSalazar2022}.  The fiducial cut is obtained by minimizing the number of particles in a narrow band around a straight line in this space, as illustrated in Figure~\ref{fig:split_accvr}.  We have slightly modified our orbiting/infall cut by moving the ``cut'' line up or down by the width of the band.  We anticipate that this range of variation is roughly comparable to the inherent systematic uncertainty in our orbiting/infall split method.

\begin{figure*}
    \centering
    \includegraphics[width=\linewidth]{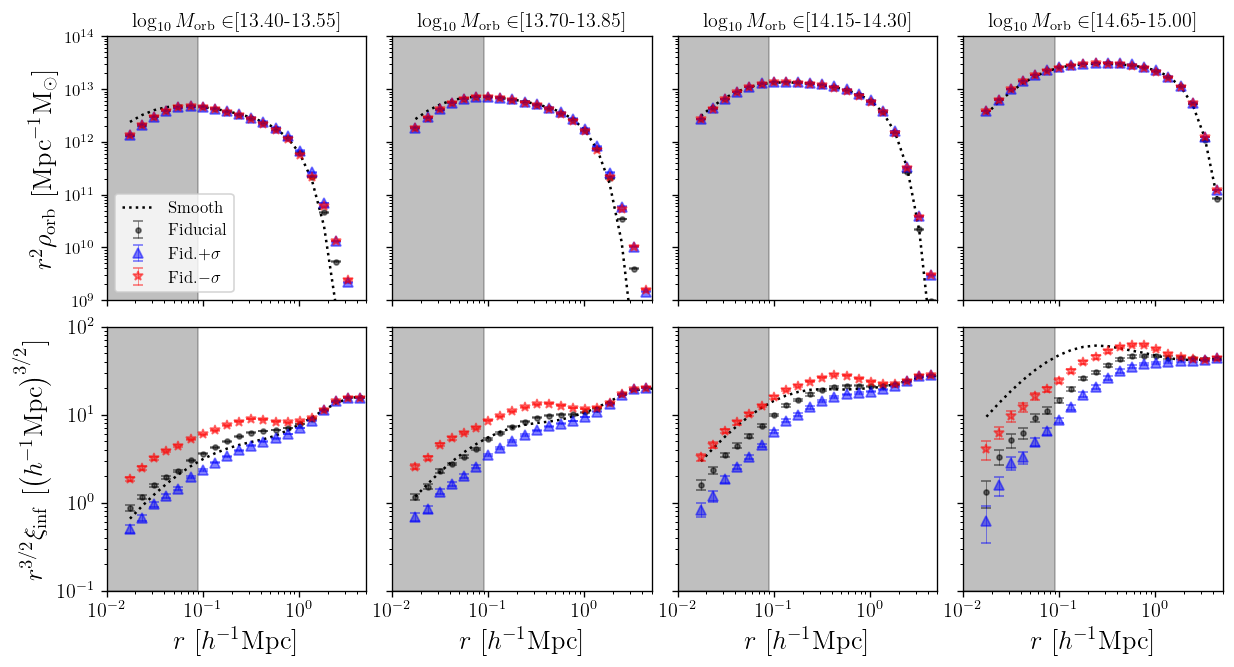}
    \caption{Orbiting density profile $\rhoorb$ and infall correlation function $\xiinf$ as determined using three different orbiting/infall splits.  Each column corresponds to a different mass bin, with mass increasing from left to right. The black data points correspond to our fiducial split; the triangular points to a $1\sigma$ upwards displacement from the fiducial line; and the star points, to a $1\sigma$ downwards displacement. The dotted line is the best fit model recovered from fitting $\xihm$ directly, and is agnostic to the particle classification scheme.}
    \label{fig:xihm_split_prediction}
\end{figure*}

Figure~\ref{fig:xihm_split_prediction} compares the orbiting and infall correlation function contributions to $\xihm$ from: 1) the best-fit model to the full halo--mass correlation function data; and 2) the three orbiting/infall particle splits (fiducial and $\pm 1\sigma$). We can see that systematic uncertainties in particle classification hardly impact the orbiting profiles.  By contrast, $\xiinf$ is clearly sensitive to the details of the orbiting/infall split.  Reassuringly, our best-fit smooth infall model to $\xihm$ is within the range of values spanned by the three orbiting/infall splits we considered except for our highest mass bin.  There is a clear trend with mass, such that high-mass haloes are better fit by an orbiting/infall split closer to the bottom of the band shown in Figure~\ref{fig:split_accvr}, whereas low-mass haloes favor a split closer to the top of the band.  Based on these results, we conclude that the differences in Figure~\ref{fig:xihm_fit_corner} are due to systematic errors in the particle classification. 

\subsection{The Halo--Mass Correlation Function of Rockstar Haloes}
\label{sec:traditional_haloes}

Throughout this work we focused on fitting the halo--mass correlation function of dynamical haloes.  However, the basic insight that particles in the vicinity of a halo split into orbiting and infalling particles is agnostic to the halo definition.  The question arises as to whether our model for the halo--mass correlation function can accurately describe traditional halo catalogues as well.  Here, we test whether our model can accurately describe the halo--mass correlation function in the original \rockstar\ halo catalog used in \citet{GarciaSalazar2022} as the basis for the dynamical halo catalogue employed in this work. The two catalogues differ in two ways: 1) our fiducial catalogue defines halo masses in terms $\Morb$ rather than $M_{200m}$, which scatters haloes between bins; and 2) we re-percolated the \rockstar\ algorithm by demanding dark matter particles can only orbit one halo at a time.  This condition removes a small fraction of the original \rockstar\ haloes from the catalogue.  For details, we refer the reader do \citet{GarciaSalazar2022}.

We assign every \rockstar\ halo a mass $M_{\rm 200m}$, and bin the haloes accordingly. We then fit the resulting halo--mass correlation function using our orbiting+infall model. Unlike for dynamical haloes, however, we do not require that the orbiting profile integrates to the halo mass $M_{\rm 200m}$, which adds an additional parameter to our fit.  Other than that, our model for \rockstar\ haloes is identical to that employed in our description of dynamical haloes, except that all mass-dependent quantities are assumed to vary with $M_{\rm 200m}$ rather than $\Morb$.

\begin{figure*}
    \centering
    \includegraphics[width=\linewidth]{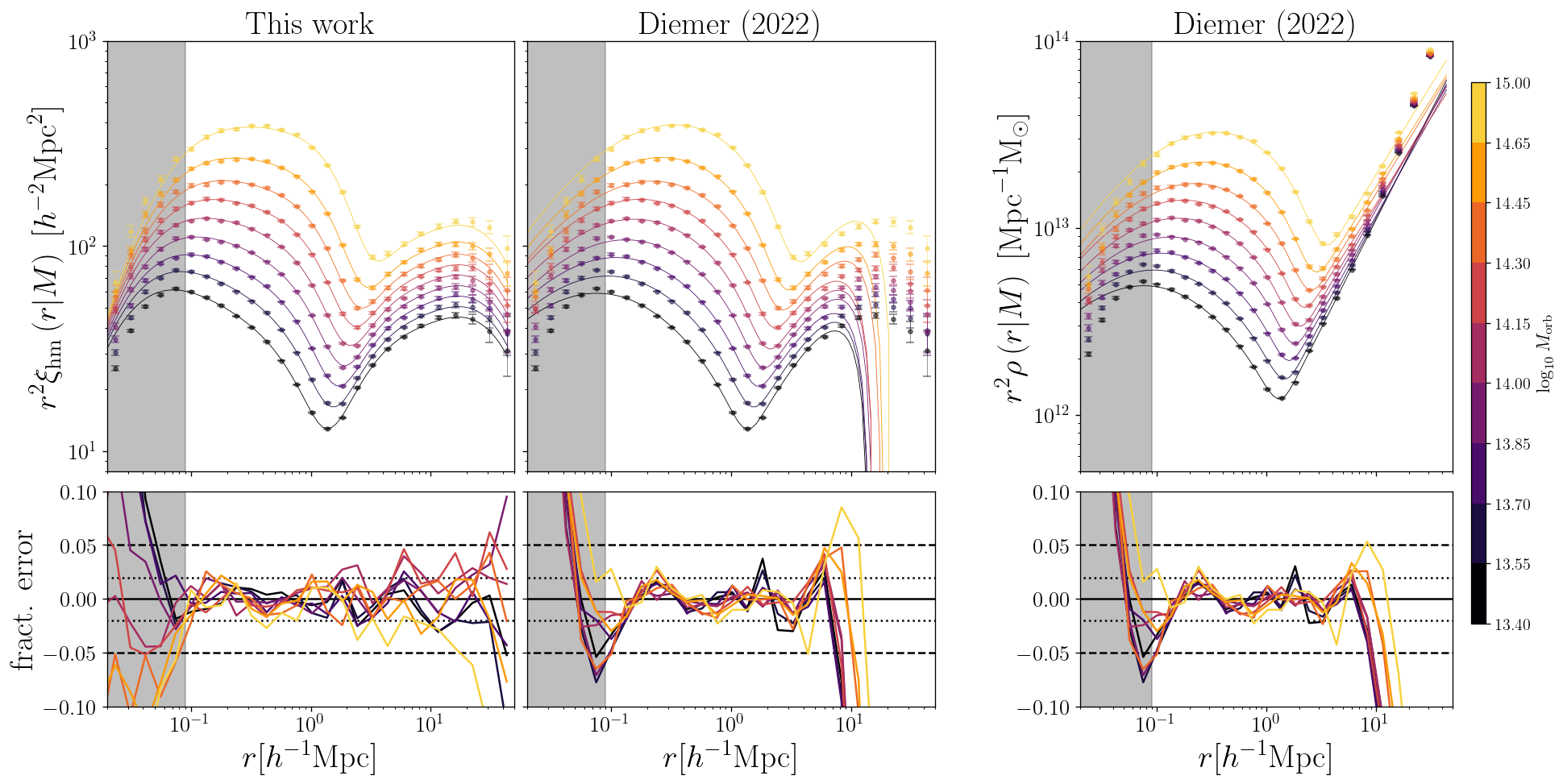}
    \caption{Halo-mass correlation function for $\textsc{Rockstar}$ haloes binned by $M_{\rm 200m}$ mass. The errors bars are jackknife. The horizontal dotted lines in the residual panels correspond to 2\% and 5\%. {\bf Left:} The best-fit model from this work.  Our model uses 9 free parameters to describe all 9 curves.
    {\bf Middle and Right panel:} Best fit models using the density profiles proposed by \citet{diemer2022b}. The latter model was designed to accurately describe the total density profile (right), but not necessarily the halo--mass correlation function (center).  Our best fit model has 9 free parameters shared across all 9 curves, while the \citet{diemer2022b} has 7 free parameters per curve.  Future work will dramatically reduce the number of free parameters in the \citet{diemer2022b} model by characterizing the evolution of the model parameters with mass.}
    \label{fig:xihm_fit_m200}
\end{figure*}

Our results are summarized in Figure~\ref{fig:xihm_fit_m200}.  The left-most panel shows the halo--mass correlation function of \rockstar\ haloes binned by $M_{\rm 200m}$, along with our best fit model.  The residuals are shown in the corresponding bottom--left panel.  Evidently, our model can accurately describe the halo--mass correlation function of \rockstar\ haloes.  The excess variance parameter $\delta$ characterising the accuracy of our model fits is $\delta=1.0\%$.

Figure~\ref{fig:xihm_fit_m200} also compares the performance of our model to the fitting functions proposed by \citet{diemer2022b}. The mass density of orbiting particles is modelled as an exponential $\rho(r) = \rho_{s}e^{S(r)}$, where:
\begin{equation}
    S(r) = -\frac{2}{\alpha}\left[\left(\frac{r}{r_{s}}\right)^{\alpha}-1\right] -\frac{1}{\beta}\left[\left(\frac{r}{r_{t}}\right)^{\beta} -\left(\frac{r_{s}}{r_{t}}\right)^{\beta}\right]
\end{equation}
and $\left\{\rho_{s}, \alpha, \beta, r_{s}, r_{t}\right\}$ are fit parameters. As discussed in \citet{diemer2022b}, we also fix $\alpha=0.18$ and fit for the other four parameters.

Their infalling model is a power law with a smooth transition given by:
\begin{equation}
    \rho(r)=\barrhom\left(\frac{\delta_{1}}{\sqrt{\left(\delta_1/\delta_{\max}\right)^2 + \left(r/R\right)^{2s}}}  -1 \right)
\end{equation}
where $\left\{\delta_1, \delta_{\max}, s\right\}$ are fit parameters, $R$ is a pivot radius chosen to be $R=R_{200m}$, and $\barrhom$ is the average mean density of the Universe.

We fit the total density field $\rho=\rhoorb+\rhoinf$ for each mass bin individually and impose the same priors as in \citet{diemer2022b} (see their Table~1). The rightmost panel in Figure~\ref{fig:xihm_fit_m200} shows our fits to the total matter density profile $\rho(r)$.  The resulting fits are of comparable quality to ours over the radial range for which the \citet{diemer2022b} model was defined ($r/R_{200m} \in [0.1,10]$).  The central panel shows this same data, but now with $r^2\xihm$ in the y-axis.  In this space the fitting formula from \citet{diemer2022b} fails at large radii.  This is expected.  \citet{diemer2022b} characterised the halo density profile, not the halo--mass correlation function. Since $\rho(r)=\bar \rho_{\rm m}(1+\xihm)$, one can achieve high accuracy in $\rho(r)$ even if the fractional accuracy in $\xihm$ is low so long as $|\xihm| \ll 1$.

\subsection{The Halo--Mass Power Spectrum}

We have constructed an accurate model for the halo--mass two-point function in configuration space.  We briefly explore the accuracy of our model in Fourier space. 
Since our model is accurate over the range of scales $r\in [0.1,50]\ \hMpc$, we test the accuracy of our model in Fourier space on scales $k\in [0.06,30]\ \hMpcinv$ (i.e., we identify $k\approx \pi/r$).  We measure the halo--mass power spectrum by pixelizing the box using a grid with $N_{\rm grid}=512$ grid points per axis.  The halo and matter fields at each pixel are calculated using a cloud-in-cell algorithm.  We then take the Fourier transform of the corresponding density fluctuations, and average over narrow shells in $k$-space.  The grid spacing limits the accuracy of this algorithm to scales larger than the Nyquist frequency $k_{\rm Ny}=\pi N_{\rm grid}/L_{\rm box}$. For our chosen grid and simulation volume, this corresponds to a relatively small $k$ value of $k_{\rm Ny}\approx 1.6\ \hMpc$.  

We overcome this limitation by using the algorithm described in \cite{Colombi2009}.  Briefly, the box is folded up on itself, thereby halving the Nyquist frequency with each fold.  For our analysis, we chose to perform five folds of the box, which increases the Nyquist frequency to $k_{\rm Ny}\approx 51.5\ (\hMpc)^{-1}$.

\begin{figure*}
    \centering
    \includegraphics[width=0.75\linewidth]{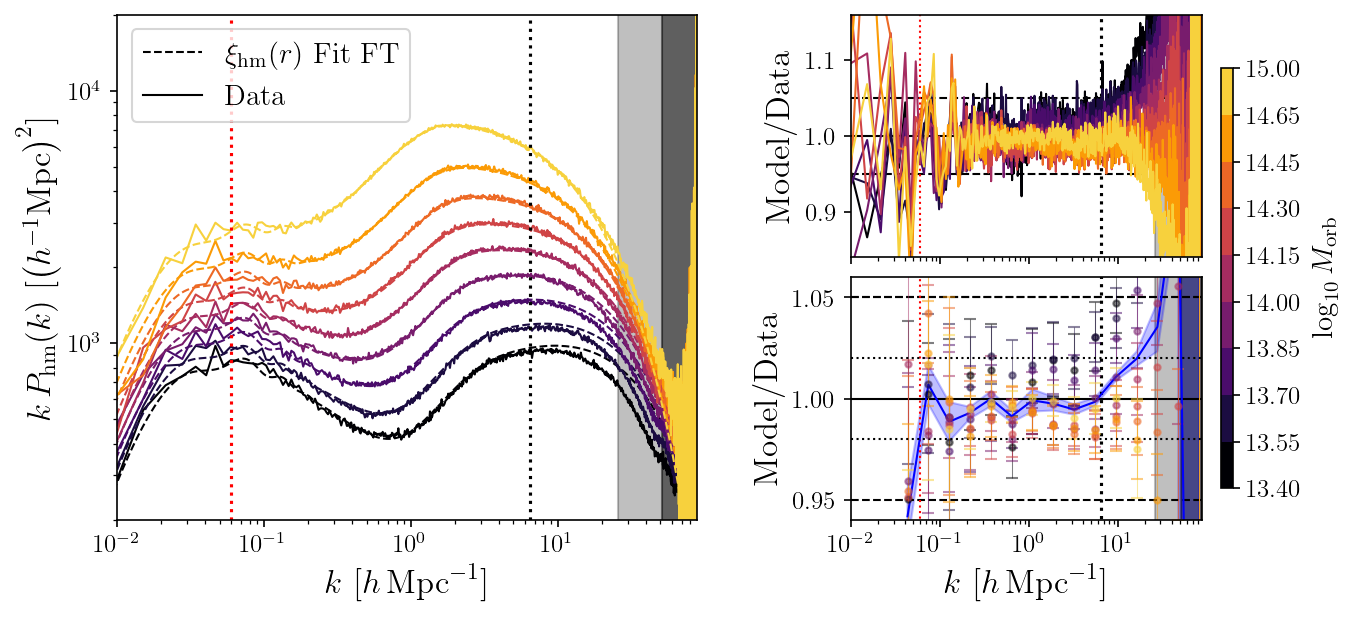}
    \caption{Halo--mass power spectrum of dynamical haloes. \textbf{Left:} The solid line is the power spectrum measured up to twice the Nyquist frequency $k_{\rm Ny}$ of the box after five folds (see text for details).  The shaded regions represent scales smaller than half the Nyquist frequency (light grey) or the Nyquist frequency (dark grey).  Each curve corresponds to a different mass bin as labelled. The dashed line is the Fourier transform of our best fit model for $\xihm(r)$ as per Table \ref{tab:parameters}. \textbf{Top-Right:} The ratio between our best fit model and the measured power spectrum. \textbf{Bottom-Right:} The ratio between the data and model fit above, averaging across relatively broad bins in $k$-space.  The blue band is the inverse-variance weighed mean offset, averaged across all mass bins.  The width of the band is the standard deviation in this ratio across all mass bins.  For reference, the horizontal dotted and dashed lines correspond to 2\% and 5\% errors respectively.
    }
    \label{fig:pkhm}
\end{figure*}

The solid lines in the left panel in Figure~\ref{fig:pkhm} show the measured power spectrum for all mass bins up to twice the Nyquist frequency of the folded box. The corresponding dashed lines show the Fourier transform of our best-fit model in configuration space. Scales smaller than the Nyquist frequency are shown as a dark grey band, while scales between half the Nyquist frequency and the Nyquist frequency are shown in light grey.    

Figure~\ref{fig:pkhm} demonstrates that there is excellent agreement between the simulation data and the Fourier transform of our model on scales $k\in[0.06,6.0]\ \hMpcinv$.  The top right panel of Figure~\ref{fig:pkhm} shows the ratio of the data to our model.  We average this ratio over comparatively broad bins in $k$-space to produce the bottom-right panel. The error bars are the error on the mean in each of these broad bins.  The blue band averages the ratios recovered in each of our different mass bins using inverse variance weights.  The width of the band is defined as the standard deviation of ratios as measured using all 9 mass bins. We see that our model for $P_{\rm hm}$ becomes highly biased at high $k$ in a mass-dependent way: while our model accurately describes high mass bins, our model overestimates the power at high $k$ for our low mass bins.  It is possible that this overestimate is due to the resolution limit of the simulation, a possibility that we will address in future work. For now, we quantify the accuracy of our model for $k\in [0.06,0.6]\ \hMpcinv$ by fitting for the excess variance required for our model to give an adequate description of the data.  When we do so, we find the best fit excess variance is $1.2\%$ at $1\sigma$. 


%% file: sections/summary.tex
\section{Summary and Discussion}
\label{sec:discussion}

We have proposed a new halo model prescription for the halo--mass correlation function based on the orbiting/infall dichotomy established in \citet{GarciaSalazar2022}.  The existence of this dichotomy implies that the 1-halo/2-halo split in our model is better physically motivated than that employed in standard halo model approaches. Our full model is given by Eq.~\eqref{eq:full_model}. The density profile $\rhoorb$ is given by Eq.~\eqref{eq:orb_model}, while the infall correlation function is given by Eq.~\eqref{eq:inf_model}. The function $\xilarge(r)$ is defined as per Eq.~\eqref{eq:large}. 

The orbiting profile $\rhoorb$ is as a running power law with index $\alpha(r)$. This running power law is exponentially truncated at large scales.  The model has three free parameters: $\rh$, $\alpha_\infty$, and $a$.  The parameter $\rh$ governs the spatial extent of the halo, and is therefore dubbed the \it halo radius. \rm  The parameter $\alpha_\infty$ is the large-scale limit of $\alpha(r)$, while the parameter $a$ governs the running of $\alpha$. We find $a$ can be modeled as being mass-independent, making $\rh$ the only radial scale associated with our halo profile.  Unsurprisingly, $\rh$ scales with mass: the more massive a halo, the bigger it is. In addition, $\alpha_\infty$ is close to isothermal, but varies slowly as a function of mass.  Note that one could use this variation in slope to define a ``scale radius’' as was done in \citet{diemer2022b}, so either parameterization is valid.

The infall correlation model assumes $\xiinf \rightarrow b\xilarge$ where $b$ is the linear halo bias, and $\xilarge$ is a suitably defined model for $\ximm$ on large scales.  After considering several options, we settled for the empirical approximation $\ximm \approx B\xizel$ where $B=0.948$ at $z=0$.  This approximation is accurate to better than $1\%$ on scales $r\in[30,50]\ \hMpc$, and extrapolates reasonably well to small scales.  We further modify the basic model $\xiinf=b\xilarge$ by incorporating: 1) a newly discovered scaling of the infall correlation function (see Figure~\ref{fig:inf_ratio}); and 2) a scale-dependent bias $\beta(r)$ that boosts the clustering amplitude around haloes on $r\lesssim 10\ \hMpc$ scales.  The latter is modeled as a cored power law, where the core scale can be associated with the halo radius $\rh$.

Our final description of the infall profile for a single mass bin has three mass-dependent parameters ($b$, $\eta$, and $\gamma$) and two population parameters $\rinf$ and $\mu$. 
The parameter $b$ is the linear halo bias. The parameter $\eta$ accounts for the newly discovered scaling of $\xiinf$ shown in Figure~\ref{fig:inf_ratio}.  Finally, the parameters $\rinf$ and $\gamma$ are the amplitude and slope of the cored power law defining the scale-dependent bias.  The core radius is set to $\mu\rh$, where $\rh$ is the halo radius. We expect that the amplitude of the small-scale dependent bias $\rinf$ must be related to nonlinear structure growth, so we anticipate $\kNL\rinf \approx 1$, where $\kNL$ is the non-linear scale defined by $\Delta_{\rm lin}^2(\kNL)=1$.  Empirically, we recovered $\kNL\rinf\approx 2/3$.

It is worth commenting on the differences between the model presented in this work and that of \citet{GarciaRozo2021} and \citet{zhouhan23}.  Both of these works defined haloes using physically motivated radial apertures.  Moreover, both works explicitly incorporated the impact of halo exclusion into their models for $\xihm$.  A consequence of this type of modelling is that the one- and two-halo terms overlap: the one-halo term extends into the two-halo regime, and vice-versa.  This is a necessary ingredient for achieving high-accuracy.  In those works, this overlap comes about in an ad-hoc way, whereas in our model it is a natural consequence of our definition.  As a specific example, the two-halo term in the model of \citet{GarciaRozo2021} does not extend as far into the one-halo regime as our infall model does.  This difference is purely cosmetic if one is agnostic to the physical interpretation of the fitting function.  However, we have demonstrated that our one and two halo terms have a clear physical interpretation that is lacking in the work of \citet{GarciaRozo2021}.

One possible source of concern is that we have made no attempt to incorporate halo-exclusion effects. However, we do not believe this is problematic.  Although the current data set is expected to be impacted by halo-exclusion effects, we anticipate these will cease to exist in the future.  When haloes are defined as collections of orbiting particles, the restriction that haloes cannot overlap vanishes: if a small halo is falling into a larger halo, and the small halo has not yet experienced its first pericentric passage, then that halo will still be considered a parent halo.  This is true \it even if the smaller halo is entirely contained within the radius of the larger halo. \rm That is, a dynamical halo finder will produce halo catalogs with no halo exclusion effects.  For this reason, we have chosen to ignore halo exclusion effects in our current model.  We will present a halo finder that produces halo catalogues with no halo exclusion effects in future work. 


%% file: sections/app_mass.tex
\section{Orbiting Mass Correction}
\label{app:mass_correction}

By definition, the orbiting density profile $\rho(r)$ integrates to the total orbiting mass of a halo. Suppose the density profile in simulations is well described by our model $\rho_{\rm model}(r)$ at scales $r>R$, whereas $\rho_{\rm data}(r)$ differs from ou rmodel on scales below $R$ (e.g. due to resolution effects). By integrating our model over all space we must account for the deficit or excess in mass due to the extrapolation of the model to small scales. We define the orbiting mass correction to be:

\begin{equation}
    \Delta\Morb = \int\limits_{r<R}\left(\rho_{\rm model} - \rho_{\rm data}\right)dV
\end{equation}

We have chosen $R$ to be six times the simulation softening length. Whenever we refer to the ``orbiting mass'' of stacked profiles in the text, it is always including the correction to the ensemble mass average as:
\begin{equation}
    \Morb=\langle\Morb\rangle + \Delta\Morb
\end{equation}
This amounts to at most a 2\% correction to the halo mass for the lowest mass haloes.
